\documentclass[preprints,article,accept,moreauthors,LaTeX, dvi2pdf]{Definitions/mdpi} 

\firstpage{1} 
\pubvolume{1}
\issuenum{1}
\articlenumber{0}
\pubyear{2021}
\copyrightyear{2020}
\datereceived{} 
\dateaccepted{} 
\datepublished{} 
\hreflink{https://doi.org/} 
\pdfoutput=1

\usepackage{slashed}
\newcommand{\beq}{\begin{equation}}
\newcommand{\eeq}{\end{equation}}
\newcommand{\bea}{\begin{eqnarray}}
\newcommand{\eea}{\end{eqnarray} }
\def\Tr{\mbox{Tr}}

\Title{A brief review of Implicit Regularization and its connection with the BPHZ theorem}

\TitleCitation{A brief review of Implicit Regularization}

\Author{Dafne Carolina Arias-Perdomo $^{1}$, Adriano Cherchiglia $^{1}$*, Brigitte Hiller$^{2}$ and Marcos Sampaio $^{1}$
}

\AuthorNames{Dafne Carolina Arias-Perdomo, Adriano Cherchiglia, Brigitte Hiller and Marcos Sampaio}

\AuthorCitation{Arias-Perdomo, D. C.; Cherchiglia,A.; Hiller H.; Sampaio, M.}

\address{%
$^{1}$ \quad CCNH, Universidade Federal do ABC, 09210-580 , Santo Andr\'e - SP, Brazil;\\
$^{2}$ \quad CFisUC, Department of Physics, University of Coimbra, P-3004-516 Coimbra, Portugal}

\corres{Correspondence: adriano.cherchiglia@ufabc.edu.br}

\abstract{Quantum Field Theory, as the keystone of particle physics, has allowed great insights to deciphering the core of Nature. Despite its striking success, by adhering to local interactions, Quantum Field Theory suffers from the appearance of divergent quantities in intermediary steps of the calculation, which encompasses the need for some regularization/renormalization prescription. As an alternative to traditional methods, based on the analytic extension of space-time dimension, frameworks that stay in the physical dimension have emerged, Implicit Regularization is one among them. We briefly review the method, aiming to illustrate how Implicit Regularization complies with the BPHZ theorem, which implies that it respects unitarity and locality to arbitrary loop order.}

\keyword{Renormalization/Regularization; Renormalization Group Functions; BPHZ theorem} 

\begin{document}


\section{Introduction}

In Feynman diagram calculations, the building blocks of the perturbative expansion, the transition amplitudes, contain apparent divergences at intermediate stages, and yet all physical quantities
we compute in perturbation theory are expected to be finite. A technical problem is immediately posed as one has to perform algebraic operations with divergent quantities. The naive solution is simply to ``regulate'' the divergences or make them ``manifestly finite'' so they can make sense at a mathematical level and hope that somehow the physical answer is meaningful
when the regularization is lifted. It is important to notice that ultraviolet (UV) and infrared (IR) divergences, in the high and low energy domain, respectively,  are unavoidable byproducts of the very construction of quantum field theoretical model as an effective theory. In building Feynman diagram amplitudes using Feynman rules, the product of two distributions is no longer a distribution and is, therefore, ill-defined in the short distance (UV) limit. It does not possess a Fourier transform, for instance \cite{FREEDMAN}. On the other hand
IR divergences afflict massless theories and are a consequence
of idealizations for the physical situation: taking the region of space-time to be infinite and supposing that massless particles can be detected with
infinitely precise energy-momentum resolution. Quantum field theoretical divergences arise in other ways, for instance through the lack
of convergence of the perturbation series, which at best is an asymptotic series. Despite all that, the standard model of particle physics is the best-tested physical theory ever. For instance, the theoretical and experimental deviation of the anomalous magnetic moment of the muon $a_\mu$  has recently been measured to be \cite{MUON}:
\begin{equation}
    \Delta a_\mu = a_\mu^{exp}-a_\mu^{th} = (25.1 \pm 5.9) \times 10^{-10}.
\end{equation}
Most calculations in the standard model including some supersymmetric extensions \footnote{ Analytical continuation in the space-time dimensions clashes with the invariance of an action with respect to supersymmetric transformations. It only holds in general for specific values of the space-time dimension due to the fact that a necessary condition for supersymmetry is equality of Bose and Fermi degrees of freedom}
are performed using the variants of dimensional regularization (DREG) although 
it is well-known that DREG has certain complications with
the definition of $\gamma_5$ matrices \footnote{In d=4 $\gamma_5=\gamma^5=i\gamma^0\gamma^1\gamma^2\gamma^3$.}, and thus the treatment of the chiral theories is subtle. In this case, we have recently argued that care must be exercised even when we apply a framework that works essentially in the physical dimension \cite{BRUQUE,VIGLIONI,JOILSON}. The essence of the dimensional regularization is to extend the space-time dimensionality slightly
away from the physical dimension and then take the physical limit after the actual calculation \cite{DREG}.

UV and IR divergences are conceptually different in the sense that, unlike IR infinities, UV divergences cannot be excused away for an idealization of a physical situation. UV infinities can however be renormalized. The renormalization program allows for unambiguously extracting numerical predictions for renormalizable quantum field theories order by order in perturbation theory by redefining the physical constants in the Lagrangian, such as masses, physical fields and coupling constants. Such a program is so successful that renormalizability has become a criterium for a sensible theory after its success in Quantum Electrodynamics. Today we know that this is a simplistic view. A more sensible way to understand is the effective approach to the problem of UV divergences in which one has to declare
explicit restrictions on the domain of energy scales of QFTs and adjust the sensitivity to high energy phenomena with the tools of renormalization theory \cite{RIVAT}. From the practical viewpoint, the problem is even more subtle. At the level of scattering cross-sections, local cancellation of infrared singularities between the so-called real and virtual emission processes in QED and QCD processes mixes UV and IR degrees of freedom in dimensional methods. Whilst finitude is guaranteed by  the Kinoshita-Lee-Nauenberg theorem \cite{KLN,Lee:1964is}
 which states that suitably defined inclusive
quantities will indeed be free of singularities in the massless limit, the physical origin of such cancellations is obscured in DREG as  UV and IR divergences can cancel each other.

DREG is a powerful
regularization technique that is convenient not only to make
calculations but also to
prove theorems to all orders in perturbation theory respecting the relevant Ward-Slavnov-Taylor identities \cite{BM}. Any alternative regularization to DREG should be consistent at arbitrary order in perturbation theory in consonance with the renormalization program of absorbing the infinities into the physical parameters of the theory.

The renormalization program is mathematically established by the Bogoliubov– Parasiuk–Hepp–Zimmermann (BPHZ) theorem \cite{BPHZ,BPHZ2}. This scheme was originally developed by Bogoliubov and Parasiuk in terms of a recursive subtraction operation, often called Bogoliubov's R-operation \cite{BPHZ}. This framework makes it possible to subtract overlapping and nested UV divergences in Feynman integrals in a consistent way with perturbation theory . In the BPHZ scheme, the renormalization constants expressed by counterterms at the Lagrangian level are identified with local counterterms at the level of the integrands associated with Feynman graphs. In principle, to render an amplitude UV finite, the BPHZ scheme can be carried out without the need for regularization. In practice, one must adopt a regularization scheme to compute physical quantities. In the presence of a  regulator, the BPHZ-scheme provides a consistent way to separate the potentially complicated finite parts of Feynman integrals from the divergent parts.
 An alternative proof for the finiteness of the renormalized Feynman Integral was given by Zimmermann, by means of the recursive Bogoliubov's R-operation which leads to a sum over forests of graphs giving rise to the  Zimmermann's forest formula \cite{BPHZ}. The latter is an elegant and comparably simpler proof for the finiteness of Feynman Integrals and works directly in momentum space. A nice review of BPHZ method can be found in \cite{HERZOG}.
 
 The purpose of this review is to show that a regularization scheme called Implicit Regularization (IREG) \cite{Battistel:1998sz}, that works entirely in the physical dimension of the model can be implemented to all orders in perturbation theory. The distinguished feature of IREG is that the UV divergences are displayed as loop integrals free of external momentum dependence. While such program is somewhat trivial at the one-loop level (as one algebraic identity at the integrand level is enough to extract basic divergent loop integrals), at higher-order in perturbation theory this program is highly non-trivial as it must comply with the BPHZ theorem. A crucial question is whether the divergences at each order in perturbation theory can be expressed in terms of loop integrals according to Bogoliubov's recursion formula.
 
 In this work, we show that IREG respects unitarity and locality in the BPHZ sense to arbitrary loop order.  The renormalization group functions can be obtained without explicit evaluation of the basic divergent integrals by means of a characteristic renormalization constant defined at one-loop level. We illustrate our framework using a scalar field theory and generalize to abelian and non-abelian theories. We verify that such program preserves symmetry evoking momentum routing invariance in the loops of Feynman diagrams, which defines a constrained scheme where surface terms are set to vanish.

\section{IREG and the BPHZ algorithm}
\label{sec:ireg}

For simplicity, in the context of this review, we will only consider massless theories and integrals that are free from infrared divergences. We will also be restricted to a space-time with $2n$ dimensions, where $n$ is an integer. 
In general, once a $N$-loop amplitude of a Feynman graph with $L$ external legs is known, 
the strategy of IREG is to remove all external momenta from UV divergences, expressing them as a linear combination of basic divergent integrals with one loop momentum only. To fulfill this objective, we are required to perform $(N-1)$ integrations, even though the order in which they must be realized is not immediately clear. In \cite{ADRIANO} we proposed a systematic procedure to categorize the order of integration which, as a byproduct, displays automatically the counterterms to be subtracted by Bogoliubov's recursion formula. In this work, we will review its main steps. In order to settle the notation, consider that the integral in $k_{l}$ is the $l$-th we are going to deal with whose general form is given by 
\begin{align}
&I^{\nu_{1}\ldots \nu_{m}}\!=\!\!\int\limits_{k_{l}}\!\frac{A^{\nu_{1}\ldots \nu_{m}}(k_{l},q_{i})}{\prod_{i}[(k_{l}-q_{i})^{2}-\mu^{2}]}\ln^{l-1}\!\left(\!-\frac{k_{l}^{2}-\mu^{2}}{\lambda^{2}}\right)\!,
\label{I}
\end{align}
\noindent
where $l=1\cdots N$, $\int_{k_{l}}\equiv\int d^{\text{2n}}k_{l}/(2 \pi)^{\text{2n}}$ (for n integer), $q_{i}$ is an element (or combination of elements) of the set $\{p_{1},\ldots,p_{L},k_{l+1},\ldots,k_{N}\}$,  and $\mu^{2}$ is an infrared regulator. 

We recall that only infrared safe amplitudes are considered, which implies that the limit $\mu^2\rightarrow 0$ is well-defined for the amplitude as a whole, to be taken as the last step of our calculation.
The symbol $\lambda$ stands for an arbitrary non-vanishing parameter 
that will play the role of the renormalization group scale in the context of IREG. 
It first appears at one-loop level, surviving to
higher-orders due to the use of Eq.~(\ref{scale}) as we discuss 
at the end of this section.
The function $A^{\nu_{1}\ldots \nu_{m}}(k_{l},q_{i})$ may contain constants and all possible combinations of $k_{l}$ and $q_{i}$ compatible with the Lorentz structure. In the context of gauge theories, it would come from derivative couplings, Dirac algebra, etc. We argue in \cite{nosso paper QCD} that the form of $A^{\nu_{1}\ldots \nu_{m}}(k_{l},q_{i})$, coming from a specific Feynman diagram, is unique after the following steps are taken:

\begin{enumerate}[label=(\Alph*)]
\item Internal symmetry group and the usual Dirac algebra must be dealt with first. As extensively discussed in \cite{BRUQUE}, identities only valid in strictly n-dimensional spaces (n integer) such as $\{\gamma_5,\gamma_\mu\}=0$ must not be used inside divergent amplitudes \cite{BRUQUE,VIGLIONI,JOILSON}. 

\item 
The requirement of numerator/denominator consistency implies that terms with internal momenta squared in the numerator must be canceled against the denominator. 
For instance,
\begin{align}
\int_{k,q}\frac{k^2}{k^2 q^2 (k-q)^2}\bigg|_{\text{ IREG}} = \int_{k,q}\frac{1}{q^2(k-q)^2}\bigg|_{\text{ IREG}},
\end{align}
where we consider n=2, $\int_k \equiv \int d^4k/(2 \pi)^4$. 
In the same vein, symmetric integration in divergent amplitudes cannot be enforced. That is, 
	\beq
	\Bigg[\int_k k^{\mu_1}\cdots k^{\mu_{2m}} f(k^2)\Bigg]^{\text{ IREG}}\neq\;\frac{g^{\{\mu_1 \mu_2 } \cdots g^{\mu_{2m-1} \mu_{2m} \}}}{(2m)!} \Bigg[\int_k k^{2m} f(k^2)\Bigg]^{\text{ IREG}},
	\label{eq:sym}
	\eeq
where  the curly brackets indicate symmetrisation over Lorentz indices.
\end{enumerate}
After these steps, the resulting multi-loop integrand can be manipulated consistently in the framework of  IREG, meaning that 1) each overall-divergent amplitude is separated into a unique finite expression plus a divergent part, 2) power-counting finite expressions are not modified, and 3) linearity under the regularization operation $R$ is preserved namely, $[aF + b G]^{R} = a[F]^{R}+b[G]^{R}$, where $F$ and $G$ are Feynman integrals, $a,\;b$ are quantities that may only depend on external momenta and/or masses, not the internal loop momenta. Therefore, they can be safely pulled out of the integral.
%
%
Moreover, the UV content of ${\cal A}_n$ will be cast in terms of well-defined basic divergent integrals, which need not to be evaluated as we will discuss soon. 


 Given that a normal form for $A^{\nu_{1}\ldots \nu_{m}}(k_{l},q_{i})$ was achieved, we apply the rules of IREG: 
 
 \begin{enumerate}[label=(\alph*)]

	\item Starting at one loop (which is equivalent to set $l=1$ in Eq.~\ref{I}), we assume an implicit regulator which allows us to remove the external momenta dependence (encoded in $p_{i}$) from the UV divergent part of the amplitude by using the identity
	\begin{align}
	\frac{1}{(k-p_{i})^2-\mu^2}=\sum_{j=0}^{n_{i}^{(k)}-1}\frac{(-1)^{j}(p_{i}^2-2p_{i} \cdot k)^{j}}{(k^2-\mu^2)^{j+1}}
	+\frac{(-1)^{n_{i}^{(k)}}(p_{i}^2-2p_{i} \cdot k)^{n_{i}^{(k)}}}{(k^2-\mu^2)^{n_{i}^{(k)}}
		\left[(k-p_{i})^2-\mu^2\right]},
	\label{ident}
	\end{align}
	in the propagators (for simplicity, we have defined $k_{1}=k$). As briefly discussed before, $\mu \rightarrow 0$ is a fictitious mass (infrared regulator). 
	It should be emphasized that, since the starting integrals are IR-safe, the infrared regulator will only be needed in intermediate steps of the calculation, canceling in the end result. Therefore, gauge invariance will not be spoiled. After the first step, we can define basic divergent integrals (BDI's) as
	\bea
	I_{\text{log}}(\mu^2)&\equiv& \int_{k} \frac{1}{(k^2-\mu^2)^{n}}
	\eea
	\bea
	I_{\text{log}}(\mu^2)&\equiv& \int_{k} \frac{1}{(k^2-\mu^2)^{n}},\quad \quad
    I_{\text{log}}^{\nu_{1} \cdots \nu_{2r}}(\mu^2)\equiv \int_k \frac{k^{\nu_1}\cdots
		k^{\nu_{2r}}}{(k^2-\mu^2)^{r+n}}, \nonumber \\
    I_{\text{quad}}(\mu^2)&\equiv& \int_k \frac{1}{(k^2-\mu^2)^{n-1}},\quad \quad
	I_{\text{quad}}^{\nu_{1} \cdots \nu_{2r}}(\mu^2)\equiv \int_k \frac{k^{\nu_1}\cdots
		k^{\nu_{2r}}}{(k^2-\mu^2)^{r+n-1}}.
	\eea

\item BDI's with Lorentz indices $\nu_{1} \cdots \nu_{2r}$ are systematically reduced to linear combinations of BDI's without Lorentz indices (with the same superficial degree of divergence) since we comply with invariance under shifts of the integration momenta and numerator-denominator consistency \cite{BRUQUE}. Therefore, the total derivatives with respect to the internal momenta must vanish, e.g.
\bea
\int_k\frac{\partial}{\partial k_{\mu}}\frac{k^{\nu}}{(k^{2}-\mu^{2})^{n}}&=&2n\Bigg[\frac{g_{\mu\nu}}{2n}I_{\text{log}}(\mu^2)-I_{\text{log}}^{\mu\nu}(\mu^2)\Bigg]=0,\label{ST1L}
\\
\int_k\frac{\partial}{\partial k_{\mu}}\frac{k^{\nu}}{(k^{2}-\mu^{2})^{n-1}}&=& \frac{(n-1)}{2}\Bigg[\frac{2}{(n-1)}g_{\mu\nu}I_{\text{quad}}(\mu^2)-I_{\text{quad}}^{\mu\nu}(\mu^2)\Bigg]=0.
\label{ST1Q}
\eea

\item After the last step, the divergent part of the amplitude will be given in terms of scalar BDI's only. However, since we still have to take the limit $\mu\rightarrow 0$, it can be noticed that they are ultraviolet and infrared divergent objects. To isolate these divergences defining a genuine ultraviolet divergent object we use the identity below 
\beq
I_{\text{log}}(\mu^2) = I_{\text{log}}(\lambda^2) + b_{2n}\ln \frac{\lambda^2}{\mu^2},\quad b_{2n}\equiv\frac{i}{(4 \pi)^n}\frac{(-1)^{n}}{\Gamma(n)},
\label{scale}
\eeq
which introduces $\lambda>0$ as an arbitrary mass scale (renormalization group scale). 
$I_{\text{quad}} (\mu^2)$ can be chosen to vanish as $\mu$ goes to zero \cite{QED2}. By adding the divergent part with the finite terms, the limit
$\mu \rightarrow 0$ is now well defined since the whole amplitude is power counting infrared convergent from the start. As we will present in our examples, the BDI will be absorbed in the renormalization constants 
\cite{CLANT} allowing renormalization functions to be obtained using
\beq
\lambda^2\frac{\partial I_{\text{log}}(\lambda^2)}{\partial \lambda^2}= -b_{2n}.
\eeq
\end{enumerate}

At higher loop-order ($l>1$ in Eq. \ref{I}), the procedure is completely analogous. The generalization of the previous formulas are: 

\begin{enumerate}[label=(\alph*)]
    
\item After applying in the propagators the identity
    \begin{align}
	\frac{1}{(k_{l}-q_{i})^2-\mu^2}=\sum_{j=0}^{n_{i}^{(k_{l})}-1}\frac{(-1)^{j}(q_{i}^2-2q_{i} \cdot k_{l})^{j}}{(k_{l}^2-\mu^2)^{j+1}}
	+\frac{(-1)^{n_{i}^{(k_{l})}}(q_{i}^2-2q_{i} \cdot k_{l})^{n_{i}^{(k_{l})}}}{(k_{l}^2-\mu^2)^{n_{i}^{(k_{l})}}
		\left[(k_{l}-q_{i})^2-\mu^2\right]},
	\label{ident2}
	\end{align}
	where $q_{i}$ is contained in the set
		$\{p_{1},\ldots,p_{L},k_{l+1},\ldots,k_{n}\}$, the UV divergent part of the amplitude is expressed as a linear combination of the objects below 
		\footnote{We have already set quadratic divergent BDI's to zero, as previously discussed.}
\begin{align}
I_{\text{log}}^{(l)}(\mu^2)&\equiv \int\limits_{k_{l}} \frac{1}{(k_{l}^2-\mu^2)^{n}}
\ln^{l-1}{\left(-\frac{k_{l}^2-\mu^2}{\lambda^2}\right)},\quad
\label{Ilogilog}\\
I_{\text{log}}^{(l)\nu_{1} \cdots \nu_{2r}}(\mu^2)&\equiv \int\limits_{k_{l}} \frac{k_{l}^{\nu_1}\cdots
	k_{l}^{\nu_{2r}}}{(k_{l}^2-\mu^2)^{r+n}}
\ln^{l-1}{\left(-\frac{k_{l}^2-\mu^2}{\lambda^2}\right)}.
\label{IlogLorentz}
\end{align}
\item As before, higher loop BDI's are reduced to scalar ones by considering vanishing the total derivatives

\begin{align}
\int_k\frac{\partial}{\partial k_{\nu_{1}}}\frac{k^{\nu_{2}}\cdots k^{\nu_{2j}}}{(k^{2}-\mu^{2})^{n+j-1}}\ln^{l-1}\Bigg[-\frac{(k^{2}-\mu^{2})}{\lambda^{2}}\Bigg]=0.
\label{tsdef}
\end{align}
For instance,
\begin{align}
&I_{\text{log}}^{(l)\,\mu
	\nu}(\mu^2)=\sum_{j=1}^{l}\left(\frac{1}{n}\right)^j\!\frac{(l-1)!}{(l-j)!}\!\left\{\frac{g^{\mu \nu}}{2}I_{\text{log}}^{(l-j+1)}(\mu^2)\right\}.
\label{identsurface1}
\end{align}
\item We notice, once again, that the BDI's as defined in the last step, are UV and IR divergent in the limit $\mu\rightarrow 0$. To define UV divergent terms only, we apply the identity

\begin{align}
I_{\text{log}}^{(l)}(\mu^2)=I_{\text{log}}^{(l)}(\lambda^2)-\frac{b_{d}}{l}\ln^{l}\left(\frac{\mu^2}{\lambda^2}\right)+  b_{d}\sum_{k=1}^{n-1}\binom{n-1}{k}\sum_{j=1}^{l-1}\frac{(-1)^k}{k^j}\frac{(l-1)!}{(l-j)!}\ln^{l-j}\left(\frac{\mu^2}{\lambda^2}\right).
\end{align}

The $\mu$-dependence will cancel in the amplitude as a whole, since it was IR-safe from the start. As already commented, BDI's can be absorbed in renormalization constants. We take the opportunity to emphasize that a minimal, mass-independent subtraction scheme in IREG amounts to absorb only $I_{\text{log}}^{(l)}(\lambda^2)$. To evaluate renormalization group constants, only derivatives of BDI's with respect to the renormalization scale $\lambda^2$ are required  \cite{PRD2012},
\begin{align}
\lambda^2\frac{\partial I_{\text{log}}^{(l)}(\lambda^2)}{\partial \lambda^{2}}&=-(l-1)\, I_{\text{log}}^{(l-1)}(\lambda^2)- b_{2n} \,\, \alpha_{2n}^{(l)}\, ,
\label{gerder}
\end{align}
where $n \ge 2$, $\alpha_{4}^{(l)} = (l-1) !$, $\alpha_{6}^{(2)} = 3/2$ (a general formula for $\alpha_{2n}^{(l)}$ can be found in \cite{PRD2012}). 
\end{enumerate}

Once the rules of IREG are settled, we return to the discussion of our initial problem: how do we identify the order in which the $(n-1)$ integrals must be performed? We will present below a systematic choice that can be done in a way to display the terms to be subtracted by Bogoliubov's recursion formula implying that the method complies with Lorentz invariance, locality, and unitarity. 

The main idea is to adapt identity (\ref{ident2}) in such a way that the UV divergent behaviour of the amplitude as the internal momenta goes to infinity in all possible ways can be clearly identified. 
For ease of the reader, we consider that $q_{i}$ will only denote external momenta ($p_{i}$) and $k$ is an arbitrary internal momentum. By using the binomial formula, $(p_i^2-2p_i \cdot k)^{j}$ can be expanded to yield
\begin{align}
\frac{1}{(k-p_i)^2-\mu^2}=\!\sum_{l=0}^{2(n_{i}^{(k)}-1)}f_{l}^{\;(k,\;p_{i})}+{\bar{f}}^{\;(k,\;p_{i})}
\label{identbphz},
\end{align}
where we defined,
\begin{align}
&f_{l}^{\;(k,\;p_{i})}\equiv\sum_{j=0}^{\left\lfloor l/2\right\rfloor}\Theta(B)\binom{l-j}{j}
\frac{(-p_i^2)^{j}(2p_i \cdot\ k)^{l-2j}}{(k^2-\mu^2)^{l+1-j}},\quad
{\bar{f}}^{\;(k,\;p_{i})}
\equiv\frac{(-1)^{n_{i}^{(k)}}(p_i^2-2p_i \cdot k)^{n_{i}^{(k)}}}{(k^2-\mu^2)^{n_{i}^{(k)}}
\left[(k-p_i)^2-\mu^2\right]},\label{fbar}\\
&\Theta(x)\equiv\left\{\begin{array}{rc}
0&\mbox{if}\quad x\leq 0\\
1 &\mbox{if}\quad x>0
\end{array}\right.\nonumber,\quad
B\equiv n_{i}^{(k)}+j-l,\quad\left\lfloor x \right\rfloor\equiv\mbox{max}\{n\in\mathcal{Z}|n\leq x\}\nonumber.
\end{align}

As can be inspected, the terms $f_{l}^{\;(k,\;p_{i})}$ behave as $k^{-(l+2)}$ when $k\rightarrow \infty$ by construction while the value of $n_{i}^{(k)}$ is chosen to guarantee the UV finitude of ${\bar{f}}^{(k,\;p_{i})}$. The above identity is the keystone of our procedure whose application to an arbitrary Feynman amplitude is summarized as follows:

\begin{enumerate}
\item Identify which propagators depend on the external momenta, then apply identity (\ref{identbphz});
\label{pro}
\item Obtain the minimum value of all $n_{j}^{(k_{i})}$ necessary to guarantee the finitude of terms that contain ${\bar{f}}^{\;(k_{i},\;p_{j})}$ as $k_{i}\rightarrow \infty$ in all possible ways;
\label{n}
\label{repeat}
\item Isolate the UV divergent terms, allowing a classification in terms of the different ways that the internal momenta approach infinity to be envisaged;
\label{identify}
\item Use the rules of IREG, encoded in steps (a)-(c), in the terms identified in step \ref{identify} according to their classification;
\label{apIR}
\item Set aside the divergent terms that contain $I_{\text{log}}^{(l)}(\lambda^{2})$ and apply the procedure again on the ones that do not.
\label{save}
\end{enumerate} 

After step \ref{save}, we will obtain two types of terms: or $I_{\text{log}}^{(l)}(\lambda^{2})$ multiplies an integral or $I_{\text{log}}^{(l)}(\lambda^{2})$ multiplies only constants and/or polynomials in the external momenta. The first set will amount to the terms to be removed by applying Bogoliubov's recursion formula while the latter set will be the typical divergence of the graph, i.e., after subtraction of subdivergences. As emphasized before, the procedure just envisaged will allow IREG to be applied in a systematic way, with the byproduct of identifying the terms to be removed by Bogoliubov's recursion formula automatically \cite{ADRIANO}.

\section{Selected Examples}

In this section, we present some selected examples, aiming to pedagogically illustrate how the renormalization procedure can be implemented in IREG up to two-loop. We begin with a scalar theory, moving to more realistic theories afterward (QED and QCD). 

\subsection{Scalar theory $\phi^{3}$}

We initiate our discussion with a very simple theory, the massless $\phi^{3}$ model defined in 6 dimensions. The choice of dimensions is justified to obtain a more interesting (renormalizable) model, in which only graphs up to three external legs are divergent \cite{Muta}. The graphs with one external leg have only quadratic divergences which, from the point of view of  IREG, could be kept as BDI's. However, they will always cancel out in multiplicatively renormalizable theories \cite{ELOY,PRD2013,QUADRATIC}, and can be promptly dismissed in massless theories. Therefore, in order to perform the renormalization of the theory up to two-loop order, we need to consider graphs with only two or three external legs corresponding to the renormalization of the propagator and the vertex functions respectively.

\begin{figure}[H]
\begin{center}
\includegraphics{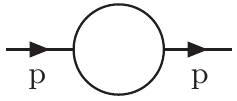}
\end{center}
\caption{Graph $P^{(1)}$}
\label{p1}
\end{figure}

We begin with the one-loop correction for the propagator depicted in fig. \ref{p1}

\noindent
whose amplitude reads\footnote{Recall we are in six dimensions which implies $\int\limits_{k}\equiv\int\frac{d^6 k}{(2 \pi)^6}$.}
\begin{align}
\Xi^{(1)}\equiv\frac{g^{2}}{2}\int\limits_{k}\!\frac{1}{k^{2}}\frac{1}{(k-p)^{2}}=\lim_{\mu^{2}\rightarrow 0}\frac{g^{2}}{2}\int\limits_{k}\!\frac{1}{(k^{2}-\mu^{2})}\frac{1}{[(k-p)^{2}-\mu^2]}\mbox{,}
\end{align}

Notice that, following the rules of IREG,  an infrared regulator was introduced in the denominators. The propagator which contains the external momentum can be rewritten in terms of $f_{l}$ and $\bar{f}$
\begin{align}
\frac{\Xi^{(1)}}{g^{2}}=\frac{1}{2}\int\limits_{k}\!\frac{1}{(k^{2}-\mu^{2})}\left[\sum_{l=0}^{2(n^{(k)}-1)}f_{l}^{\;(k,\;p)}+{\bar{f}}^{\;(k,\;p)}\right],
\end{align}
while $n^{(k)}$ is chosen in order to assure the finitude of the term containing ${\bar{f}}^{\;(k,\;p)}$. In this specific example, we find by power counting that $n^{(k)}>2$, adopting $n^{(k)}=3$. The divergent terms can be promptly identified by remembering  that $f_{l}^{\;(k,\;p)}$ behaves like $k^{-(l+2)}$ 

\begin{enumerate}
\item{Quadratic divergence
\begin{align}
\int\limits_{k}\frac{f_{0}^{\;(k,\;p)}}{(k^{2}-\mu^{2})}=\int\limits_{k}\frac{1}{(k^{2}-\mu^{2})^{2}},
\end{align}}
\item{Linear divergence
\begin{align}
\int\limits_{k}\frac{f_{1}^{\;(k,\;p)}}{(k^{2}-\mu^{2})}=\int\limits_{k}\frac{2p\cdot k}{(k^{2}-\mu^{2})^{3}},
\end{align}}
\item{Logarithmic divergence
\begin{align}
\int\limits_{k}\frac{f_{2}^{\;(k,\;p)}}{(k^{2}-\mu^{2})}=\int\limits_{k}\frac{1}{(k^{2}-\mu^{2})^{3}}\left[\frac{(2p\cdot k)^{2}}{(k^{2}-\mu^{2})}-p^{2}\right]=-\frac{p^{2}}{3}I_{\text{log}}(\mu^{2}).
\end{align}}
\end{enumerate}

For pedagogical reasons we showed the quadratic and linear divergences, even though they vanish in the limit $\mu^{2}\rightarrow 
0$. 
The remaining (UV finite) terms amount to
\begin{align}
\frac{1}{2}\int\limits_{k}\frac{f_{3}^{\;(k,\;p)}+f_{4}^{\;(k,\;p)}+{\bar{f}}^{\;(k,\;p)}}{(k^{2}-\mu^{2})}&=\frac{1}{2}\int\limits_{k}\frac{1}{(k^{2}-\mu^{2})^{4}}\left[-4p^2(p\cdot k)+p^{4}
-\!\frac{(p^2-2p \cdot k)^{3}}{(k-p)^2-\mu^2}\!\right]\nonumber\\
&=\frac{p^{2}b_{6}}{6}\ln\left(-\frac{p^{2}}{\mu^{2}}\right)-\frac{4p^{2}b_{6}}{9}+O(\mu^2).
\end{align}

After using Eq.~(\ref{scale}), the limit $\mu^{2}\rightarrow 0$ is well-defined, and we finally obtain
\begin{align}
\Xi^{(1)}=-\frac{g^2p^2}{6}\left[I_{\text{log}}(\lambda^2)-b_{6}\ln\left(-\frac{p^2}{\lambda^2}\right)+\frac{8b_{6}}{3}
\right].
\label{proP1}
\end{align}

\begin{figure}[H]
\begin{center}
\includegraphics{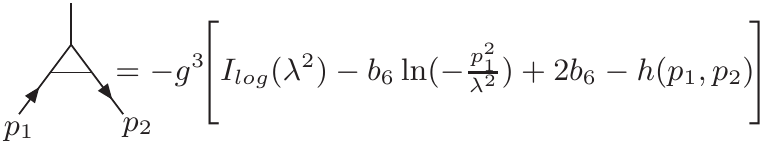}
\end{center}
\caption{Graph $V^{(1)}$}
\label{v1}
\end{figure}

Similarly we obtain the amplitude of the one-loop correction for the vertex function shown in fig. \ref{v1}

\noindent
In the amplitude corresponding to $V^{(1)}$, $h(p_{1},p_{2})$ is a function of $p_{1}$ and $p_{2}$ vanishing for $p_{2}=0$.

\subsubsection{Two loops: self-energy diagrams}

\begin{figure}[h!]
\begin{center} 
\includegraphics{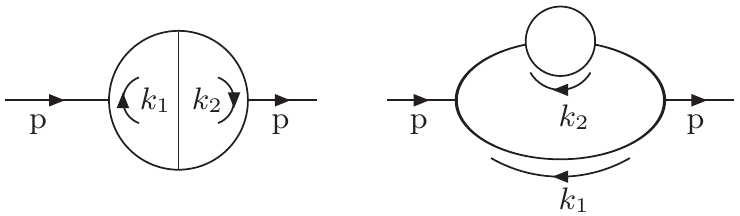}
\end{center}
\vspace{-0.35cm}
\caption{Graphs $P_{A}^{(2)}$ and $P_{B}^{(2)}$ respectively}
\label{p2}
\end{figure}

After setting the stage with the one-loop graphs, we move to the two-loop contributions. We recall that our main aim is to perform the renormalization of the theory, which implies that only the divergent parts will be kept. Starting with the scalar propagator, the diagrams needed are given in fig. \ref{p2}.

The amplitude corresponding to $P_{A}^{(2)}$ is given by
\begin{align}
\frac{\Xi_{A}^{(2)}}{ig^4}=\frac{1}{2}\int\limits_{k_{1}k_{2}}\Delta(k_{1})\Delta(k_{1}-p)\Delta(k_{1}-k_{2})\Delta(k_{2})\Delta(k_{2}-p),\quad\mbox{where}\quad\Delta(k_{i})\equiv\frac{1}{k_{i}^{2}-\mu^2}\label{eq:XA}.
\end{align}

As in the one-loop case, we start by rewriting the propagators that depend on the external momenta
\begin{align}
\int\limits_{k_{1}k_{2}}\Delta(k_{1})\Delta(k_{1}-k_{2})\Delta(k_{2})\left[\sum_{l=0}^{2(n^{(k_{1})}-1)}f_{l}^{\;(k_{1},\;p)}+{\bar{f}}^{\;(k_{1},\;p)}\right]\left[\sum_{m=0}^{2(n^{(k_{2})}-1)}f_{m}^{\;(k_{2},\;p)}+{\bar{f}}^{\;(k_{2},\;p)}\right].
\label{intid}
\end{align}

This time we have two $\;n^{(k_{i})}$ to be determined, which are chosen to guarantee the finitude of terms containing ${\bar{f}}^{(k_{i},\;p)}$ as $k_{i}\rightarrow \infty$ in all possible ways. We focus first on $\;n^{(k_{1})}$. The terms that contain ${\bar{f}}^{(k_{1},\;p)}$ can be compactly written as
\begin{align}
\int\limits_{k_{1}k_{2}}\Delta(k_{1})\Delta(k_{1}-k_{2})\Delta(k_{2}){\bar{f}}^{(k_{1},\;p)}\Delta(k_{2}-p).
\end{align} 

We want to assure the finitude of the above integral as $k_{1}\rightarrow \infty$. Two cases must be considered:
\begin{enumerate}
\item {Finitude as $k_{1}\rightarrow \infty$ and $k_{2}$ fixed: $n^{(k_{1})}>0$,}
\item {Finitude as $k_{1}\rightarrow \infty$ and $k_{2}\rightarrow \infty$: $n^{(k_{1})}>\!2$,}
\end{enumerate}
\noindent
which allows us to conclude that $n^{(k_{1})}$ should be at least 3. Similarly, we obtain $n^{(k_{2})}=3$. 

Once the values of $n^{(k_{i})}$ are known, we aim to identify the divergent terms contained in (\ref{intid}) as $k_{1}$ and/or $k_{2}$ go to infinity. There are three possibilities. We begin analysing the case $k_{1}\rightarrow \infty$ and $k_{2}$ fixed which contain divergence terms of the type
\begin{align} \int\limits_{k_{1}k_{2}}\Delta(k_{1})\Delta(k_{2})\Delta(k_{1}-k_{2})f_{l}^{\;(k_{1},\;p)}\left[\sum_{m=0}^{4}f_{m}^{\;(k_{2},\;p)}+{\bar{f}}^{\;(k_{2},\;p)}\right].
\end{align}

Since $f_{l}^{\;(k_{1},\;p)}$ goes like $k_{1}^{-(l+2)}$, we find by power counting that the divergent terms are given by
\begin{align}
A_{1}^{\Xi}&\equiv\int\limits_{k_{1}k_{2}}\Delta(k_{1})\Delta(k_{2})\Delta(k_{1}-k_{2})f_{0}^{\;(k_{1},\;p)}\left[\sum_{m=0}^{4}f_{m}^{\;(k_{2},\;p)}+{\bar{f}}^{\;(k_{2},\;p)}\right]\nonumber\\&=\int\limits_{k_{1}k_{2}}\Delta^{2}(k_{1})\Delta(k_{1}-k_{2})\Delta(k_{2})\Delta(k_{2}-p).
\label{1k}
\end{align}

In a similar fashion, the case $k_{2}\rightarrow \infty$ and $k_{1}$ amounts to
\begin{align}
A_{2}^{\Xi}&\equiv\int\limits_{k_{1}k_{2}}\Delta(k_{1})\Delta(k_{2})\Delta(k_{1}-k_{2})f_{0}^{\;(k_{2},\;p)}\left[\sum_{l=0}^{4}f_{l}^{\;(k_{1},\;p)}+{\bar{f}}^{\;(k_{1},\;p)}\right]\nonumber\\&
=\int\limits_{k_{1}k_{2}}\Delta^{2}(k_{2})\Delta(k_{1}-k_{2})\Delta(k_{1})\Delta(k_{1}-p).
\label{2k}
\end{align}

Finally we consider $k_{1}\rightarrow \infty$ and $k_{2}\rightarrow \infty$ simultaneously. The choice of $n^{(k_{i})}=3$ ($i=1,2$) guarantees us that the divergent terms must be of the type
\begin{align} \int\limits_{k_{1}k_{2}}\Delta(k_{1})\Delta(k_{2})\Delta(k_{1}-k_{2})f_{l}^{\;(k_{1},\;p)}f_{m}^{\;(k_{2},\;p)}.
\end{align}

Once again, by power counting, we obtain that $l$ and $m$ are constrained by $l+m\le2$. Cases $l=0$ and $m=0,1,2$ are already contained in $A_{1}^{\Xi}$ (Eq. \ref{1k}) while cases $m=0$ and $l=0,1,2$ are part of $A_{2}^{\Xi}$ (Eq. \ref{2k}). Thus, we are left only with the case $l=m=1$ 
\begin{align}
A_{3}^{\Xi}&\equiv\int\limits_{k_{1}k_{2}}\Delta(k_{2})\Delta(k_{1}-k_{2})\Delta(k_{1})f_{1}^{\;(k_{1},\;p)}f_{1}^{\;(k_{2},\;p)}\nonumber\\
&=\int\limits_{k_{1}k_{2}}\Delta^{3}(k_{1})\Delta(k_{1}-k_{2})\Delta^{3}(k_{2})(2p \cdot k_{1})(2p \cdot k_{2}).
\end{align}

In summary, the divergent terms are:

\begin{enumerate}
\item Case $k_{1}\rightarrow \infty$ and $k_{2}$ is fixed
\begin{align}
A_{1}^{\Xi}\!=\!\int\limits_{k_{1}k_{2}}\!\!\Delta^{2}(k_{1})\Delta(k_{1}-k_{2})\Delta(k_{2})\Delta(k_{2}-p),
\label{k1}
\end{align}
\item Case $k_{2}\rightarrow \infty$ and $k_{1}$ is fixed
\begin{align}
A_{2}^{\Xi}\!=\!\int\limits_{k_{1}k_{2}}\!\!\Delta^{2}(k_{2})\Delta(k_{1}-k_{2})\Delta(k_{1})\Delta(k_{1}-p),
\label{k2}
\end{align}
\item Case $k_{1}\rightarrow \infty$ and $k_{2}\rightarrow \infty$ simultaneously
\begin{align}
A_{3}^{\Xi}=\!\!\int\limits_{k_{1}k_{2}}\!\!\Delta^{3}(k_{1})\Delta(k_{1}-k_{2})\Delta^{3}(k_{2})(2p \cdot k_{1})(2p \cdot k_{2}).
\label{k1k2}
\end{align}
\end{enumerate}

Thus, the divergent content of $\Xi_A^{(2)}$ amounts to $A_{1}^{\Xi}+A_{2}^{\Xi}+A_{3}^{\Xi}-A_{4}^{\Xi}$. The last term corresponds to the case ($l=m=0$) 
\begin{align}
A_{4}^{\Xi}\equiv&\int\limits_{k_{1}k_{2}}\Delta(k_{2})\Delta(k_{1}-k_{2})\Delta(k_{1})f_{0}^{\;(k_{1},\;p)}f_{0}^{\;(k_{2},\;p)}=\int\limits_{k_{1}k_{2}}\Delta^{2}(k_{2})\Delta(k_{1}-k_{2})\Delta^{2}(k_{1}),
\label{k0}
\end{align}
\noindent
which must be subtracted since it was counted twice.

The above classification of the divergent terms in different cases (we consider $A_{4}^{\Xi}$ as the intersection between the $k_{1}\rightarrow \infty$ and $k_2$ fixed, $k_{2}\rightarrow \infty$ and $k_1$ fixed) is crucial to implement IREG to multi-loop Feynman graphs in a systematic way, since it gives a natural order in which the integrals must be performed. For instance, to evaluate $A_{1}^{\Xi}$ further we must perform the integral in $k_{1}$ first. For terms like $A_{3}^{\Xi}$, which are symmetric under $k_{1}\leftrightarrow k_{2}$, one may perform any of the integrals first. We should also emphasize that, 
as a byproduct, this classification will display the terms to be subtracted by Bogoliubov's recursion formula automatically as we will show. 

Returning to the divergent terms we classified, one may notice that $A_{1}^{\Xi}$ and $A_{2}^{\Xi}$ have the same structure, and the integral to be dealt with first is
\begin{align}
\int\limits_{k_{i}}\Delta^{2}(k_{i})\Delta(k_{i}-k_{j}),\quad i,j=1,2\!\!\! \quad\mbox{and}\!\!\!\quad i\neq j
\label{vertex}
\end{align}
 
It is the same amplitude of graph $V_1$ (fig \ref{v1}) by identifying $p_{1}\rightarrow k_{j}$ and setting $p_{2}=0$. Therefore we can write
\begin{align}
A_{i}^{\Xi}&=\bar{A}_{i}^{\Xi}+\alpha_{i}^{\Xi},\quad\quad i,j=1,2 \quad\mbox{and}\quad i\neq j\nonumber\\
\bar{A}_{i}^{\Xi}&\equiv\int\limits_{k_{j}}\Delta(k_{j})\Delta(k_{j}-p)\left[I_{\text{log}}(\lambda^2)\right],\nonumber\\\alpha_{i}^{\Xi}&\equiv b_{6}\int\limits_{k_{j}}\Delta(k_{j})\Delta(k_{j}-p)\left[2-\ln\left(-\frac{k_{j}^{2}-\mu^2}{\lambda^2}\right)\right].
\label{k1bphz}
\end{align}

We turn to $A_{3}^{\Xi}$. We choose to perform the integral in $k_{1}$ first (which is finite), insert the result in the integral in $k_{2}$ and use the rules of IREG to obtaining
\begin{align}
\bar{\alpha}_{3}^{\Xi}\equiv A_{3}^{\Xi}=b_{6}p^{2}\left[\frac{I_{\text{log}}(\lambda^2)}{3}
\right].
\label{resultk1k2}
\end{align}
\noindent
Similarly
\begin{align}
A_{4}^{\Xi}&=\int\limits_{k_{2}}\Delta^{2}(k_{2})\left[I_{\text{log}}(\lambda^2)-b_{6}\!\ln\left(\!\!-\frac{\!k_{2}^{2}-\!\mu^2}{\lambda^2}\right)\!\!+2b_{6}\!\right]=0
\end{align}
\noindent
in the limit $\mu^{2}\rightarrow 0$.

We will verify that the terms $\bar{A}_{i}^{\Xi}$ ($i=1,2$) are exactly the ones which are going to be subtracted by applying Bogoliubov's recursion formula. We set them aside for now and evaluate the rest ($\alpha_{i}^{\Xi}$). As usual, after using identity (\ref{identbphz}) in the propagator that depends on the external momentum, we are able to identify the divergent terms. After taking the limit $\mu^{2}\rightarrow0$, the only one that survives is 
\begin{align}
\bar{\alpha}_{i}^{\Xi}\equiv\int\limits_{k_{j}}\Delta(k_{j})f_{2}^{\;(k_{j},\;p)}\!\!\left[-b_{6}\ln\left(-\frac{k_{j}^{2}-\mu^2}{\lambda^2}\right)+2b_{6}\right]=b_{6}p^{2}\Bigg[\frac{I_{\text{log}}^{(2)}(\lambda^2)}{3}-\frac{8}{9}I_{\text{log}}(\lambda^2)
\Bigg].
\end{align}

Hence, the divergent content of $\Xi_A^{(2)}$ amounts to
\begin{align}
\frac{\Xi_{A}^{(2)\infty}}{ig^{4}}\equiv\frac{1}{2}\big(\bar{\alpha}_{1}^{\Xi}+\bar{\alpha}_{2}^{\Xi}+\bar{\alpha}_{3}^{\Xi}+\bar{A}_{1}^{\Xi}+\bar{A}_{2}^{\Xi}\big).
\label{divpa2}
\end{align}

\begin{figure}[h!]
\begin{center} 
\includegraphics{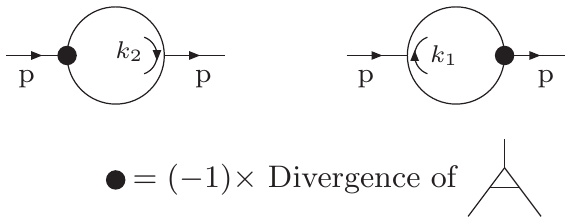}
\end{center}
\caption{Counterterms for $P_{A}^{(2)}$}
\label{cp2}
\end{figure}

As stated before, the two last terms are exactly the ones that are going to be subtracted after applying Bogoliubov's recursion formula. Explicitly, the subdivergences of this particular graph are subtracted by the counterterms shown in fig. \ref{cp2}

\noindent
whose amplitudes are, respectively
\begin{align}
\frac{ig^{4}}{2}\int\limits_{k_{2}}\Delta(k_{2})\Delta(k_{2}-p)\left[-I_{\text{log}}(\lambda^2)\right]=\frac{ig^{4}}{2}\left(-\bar{A}_{1}^{\Xi}\right),
\end{align}
\begin{align}
\frac{ig^{4}}{2}\int\limits_{k_{1}}\Delta(k_{1})\Delta(k_{1}-p)\left[-I_{\text{log}}(\lambda^2)\right]=\frac{ig^{4}}{2}\left(-\bar{A}_{2}^{\Xi}\right).
\end{align}
Notice that we are adopting a minimal subtraction scheme which, in the context of IREG, corresponds to the subtraction of basic divergent integrals \cite{Sampaio:2002ii}.
Therefore, after the subtraction of subdivergences we finally obtain
\begin{align}
\frac{\bar{\Xi}_A^{(2)}}{ig^4}\equiv\frac{b_{6}p^2}{6}\Bigg[&2I_{\text{log}}^{(2)}(\lambda^2)-\frac{13}{3}I_{\text{log}}(\lambda^2)+
\mbox{finite}\Bigg].
\label{proP2}
\end{align}

We consider next the two loop nested graph ($P_{B}^{(2)}$) whose amplitude is
\begin{align}
\frac{\Xi_B^{(2)}}{ig^{4}}=\frac{1}{2}\int\limits_{k_{1}k_{2}}\Delta^{2}(k_{1})\Delta(k_{1}-p)\Delta(k_{2})\Delta(k_{1}-k_{2}).
\end{align}

Once again, identity (\ref{identbphz}) is applied in the propagators that depend on the external momentum, allowing us to choose $n^{(k_{1})}=3$ to assure that terms containing ${\bar{f}}^{(k_{1},\;p)}$ are finite as $k_{1}\rightarrow \infty$ in all possible ways. We proceed to classify the divergent terms. It is easy to see that the case $k_{1}\rightarrow \infty$ and $k_{2}$ fixed does not have any divergent term while the case $k_{2}\rightarrow \infty$ and $k_{1}$ fixed does given below
\begin{align}
\int\limits_{k_{1}k_{2}}\Delta^{2}(k_{1})\Delta(k_{1}-k_{2})\Delta(k_{2})\left[\sum_{l=0}^{4}f_{l}^{\;(k_{1},\;p)}+{\bar{f}}^{\;(k_{1},\;p)}\right]=\int\limits_{k_{1}k_{2}}\Delta^{2}(k_{1})\Delta(k_{1}-k_{2})\Delta(k_{2})\Delta(k_{1}-p).
\end{align}

For definiteness, we denote the above integral $B_{1}^{\Xi}$. It is the only term that we have to deal with (the divergent terms from the case $k_{1}\rightarrow \infty$ and $k_{2}\rightarrow \infty$ simultaneously are contained in the above integral). It should be noticed that, although it is the original amplitude, we have now a natural order to implement IREG. After a straightforward use of the rules in the integral in $k_{2}$ we have
\begin{align}
&B_{1}^{\Xi}=\bar{B}_{1}^{\Xi}+\beta_{1}^{\Xi},\nonumber\\
&\bar{B}_{1}^{\Xi}\equiv\int\limits_{k_{1}}\Delta(k_{1})\Delta(k_{1}-p)\left[-\frac{I_{\text{log}}}{3}(\lambda^2)\right],\nonumber\\
&\beta_{1}^{\Xi}\equiv \frac{b_{6}}{3}\int\limits_{k_{1}}\Delta(k_{1})\Delta(k_{1}-p)\Bigg[\ln\left(-\frac{k_{1}^{2}-\mu^{2}}{\lambda^2}\right)-\frac{8}{3}\Bigg]
\end{align}

BY applying the procedure again in $\beta_{1}^{\Xi}$ we can obtain the following divergent terms
\begin{align}
\bar{\beta}_{1}^{\Xi}\equiv&\frac{b_{6}}{3}\int\limits_{k_{1}}\Delta(k_{1})\left[\sum_{l=0}^{2}f_{l}^{\;(k_{2},\;p)}\right]\left[\ln\left(-\frac{k_{1}^{2}-\mu^{2}}{\lambda^2}\right)\right]-\frac{8}{3}\Bigg]
=
-\frac{b_{6}p^2}{9}\Bigg[I_{\text{log}}^{(2)}(\lambda^2)-\frac{10}{3}I_{\text{log}}(\lambda^2)
\Bigg].
\end{align}


Thus, the divergent content of $\Xi_B^{(2)}$ is given by 
\begin{align}
\frac{\Xi_B^{(2)\infty}}{ig^{4}}\equiv\frac{1}{2}\left(\bar{\beta}_{1}^{\Xi}+\bar{B}_{1}^{\Xi}\right).
\end{align}

\begin{figure}[ht]
\begin{center} 
\includegraphics{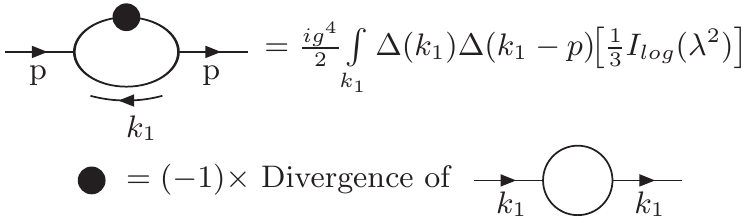}
\end{center}
\caption{Counterterm for $P_{B}^{(2)}$}
\label{c2p2}
\end{figure}

As before, the last term is exactly the the one to be removed by an application of Bogoliubov's recursion formula since the counterterm we consider in this case is shown in fig. \ref{c2p2}

After removing the subdivergence we finally obtain
\begin{align}
&\frac{\bar{\Xi}_B^{(2)}}{ig^{4}}\equiv\frac{b_{6}p^2}{18}\Bigg[-I_{\text{log}}^{(2)}(\lambda^2)+\frac{10}{3}I_{\text{log}}(\lambda^2)
+\mbox{finite}\Bigg].
\end{align}


Finally, we are able to write down the divergent part of the two point function 
at two loop order
\begin{align}
\bar{\Xi}^{(2)}_{\text{div}}\equiv\left(\bar{\Xi}_A^{(2)}+\bar{\Xi}_B^{(2)}\right)_{\text{div}}=ig^4\frac{p^2}{6}\Bigg[\frac{5b_{6}}{3}I_{\text{log}}^{(2)}(\lambda^2)-\frac{29b_{6}}{9}I_{\text{log}}(\lambda^2)
\Bigg].
\label{prop}
\end{align}

\subsubsection{Two-loop vertex renormalization}

\begin{figure}[ht]
\begin{center} 
\includegraphics{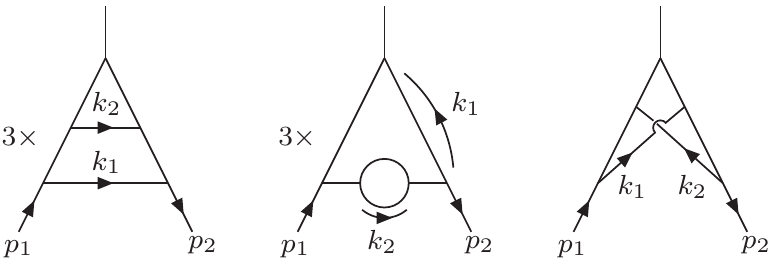}
\end{center}
\caption{Graphs $V_{A}^{(2)}$, $V_{B}^{(2)}$ and $V_{C}^{(2)}$ respectively}
\label{V2}
\end{figure}

We consider now the renormalization of the vertex. The Feynman diagrams to be evaluated are depicted in fig. \ref{V2}
The amplitude corresponding to $V_{A}^{(2)}$ is given by
\begin{align}
\frac{\Lambda_{A}^{(2)}}{-ig^{5}}\equiv\int\limits_{k_{1}k_{2}}\Delta(k_{1})\Delta(k_{2}-k_{1})\prod_{i=1}^{2}\Delta(k_{i}-p_{1})\Delta(k_{i}-p_{2}).
\end{align}

Our first task is to obtain $n^{(k_{i})}_{j}$. We apply the same procedure as before, choosing
$n^{(k_{1})}_{1}=n^{(k_{1})}_{2}=n^{(k_{2})}_{1}=n^{(k_{2})}_{2}=1$ which amounts to
\begin{align}
\int\limits_{k_{1}k_{2}}\Delta(k_{1})\Delta(k_{2}-k_{1})&\left[f_{0}^{\;(k_{1},\;p_{1})}+{\bar{f}}^{\;(k_{1},\;p_{1})}\right]\left[f_{0}^{\;(k_{1},\;p_{2})}+{\bar{f}}^{\;(k_{1},\;p_{2})}\right]\times\nonumber\\&\left[f_{0}^{\;(k_{2},\;p_{1})}+{\bar{f}}^{\;(k_{2},\;p_{1})}\right]\left[f_{0}^{\;(k_{2},\;p_{2})}+{\bar{f}}^{\;(k_{2},\;p_{2})}\right].
\end{align}

The divergent terms come only from the case $k_{2}\rightarrow \infty$ and $k_{1}$ fixed, yielding
\begin{align}
A_{1}^{\Lambda}&\equiv\int\limits_{k_{1}k_{2}}\Delta(k_{1})\Delta(k_{2}-k_{1})f_{0}^{\;(k_{2},\;p_{1})}f_{0}^{\;(k_{2},\;p_{2})}
\left[f_{0}^{\;(k_{1},\;p_{1})}+{\bar{f}}^{\;(k_{1},\;p_{1})}\right]\left[f_{0}^{\;(k_{1},\;p_{2})}+{\bar{f}}^{\;(k_{1},\;p_{2})}\right]\nonumber\\&=\int\limits_{k_{1}k_{2}}\Delta(k_{1})\Delta(k_{2}-k_{1})\Delta^2(k_{2})\Delta(k_{1}-p_{1})\Delta(k_{1}-p_{2})\nonumber\\
&=\bar{A}_{1}^{\Lambda}+\alpha_{1}^{\Lambda},\nonumber\\
\bar{A}_{1}^{\Lambda}&\equiv\int\limits_{k_{1}}\Delta(k_{1})\Delta(k_{1}-p_{1})\Delta(k_{1}-p_{2})\left[I_{\text{log}}(\lambda^2)\right],\nonumber\\
\alpha_{1}^{\Lambda}&\equiv b_{6} \int\limits_{k_{1}}\Delta(k_{1})\Delta(k_{1}-p_{1})\Delta(k_{1}-p_{2})\left[2-\ln\left(-\frac{k_{1}^{2}-\mu^2}{\lambda^2}\right)\right].
\label{vertexk1}
\end{align}

By applying the procedure in $\alpha_{1}^{\Lambda}$ we obtain the the divergent term below
\begin{align}
\bar{\alpha}_{1}^{\Lambda}&\equiv b_{6}\int\limits_{k_{1}}\Delta(k_{1})f_{0}^{\;(k_{1},\;p_{1})}f_{0}^{\;(k_{1},\;p_{2})}
\left[2-\ln\left(-\frac{k_{1}^{2}-\mu^2}{\lambda^2}\right)\right]\nonumber\\&=2b_{6}I_{\text{log}}(\lambda^2)-b_{6}I_{\text{log}}^{(2)}(\lambda^2).
\end{align}

\begin{figure}[ht]
\begin{center} 
\includegraphics{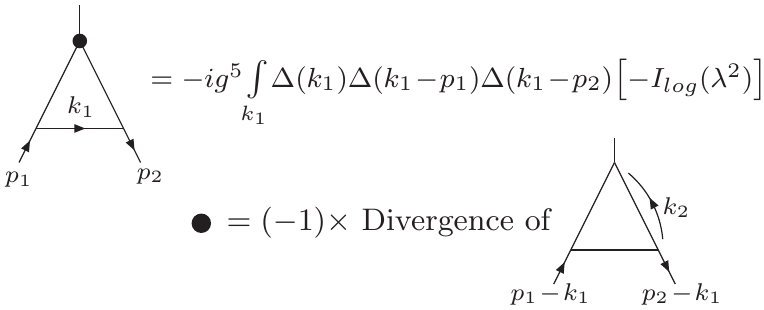}
\end{center}
\caption{Counterterm for $V_{A}^{(2)}$}
\label{c2v2}
\end{figure}

Hence, the divergent content of $\Lambda_{A}^{(2)}$ is given by
\begin{align}
\Lambda_{A}^{(2)}\Big|_{\text{div}}\equiv-ig^{5}\left[\bar{\alpha}_{1}^{\Lambda}+\bar{A}_{1}^{\Lambda}\right]
\end{align}
\noindent
where the last term is removed by adding the counterterm shown in fig. \ref{c2v2}

Therefore, after the subtraction of the subdivergence we have
\begin{align}
\frac{\bar{\Lambda}_{A}^{(2)}}{-ig^{5}}\equiv b_{6}\left[-I_{\text{log}}^{(2)}(\lambda^2)+2I_{\text{log}}(\lambda^2)+\mbox{finite}\right].
\end{align}

We move to graph $V_{B}^{(2)}$ whose amplitude is 
\begin{align}
\frac{\Lambda_{B}^{(2)}}{-ig^{5}}\equiv\frac{1}{2}\int\limits_{k_{1}k_{2}}\Delta^{2}(k_{1})&\Delta(k_{1}-p_{1})\Delta(k_{1}-p_{2})
\Delta(k_{2})\Delta(k_{2}-k_{1}).
\end{align}

After choosing 
$n^{(k_{1})}_{1}=n^{(k_{1})}_{2}=1$, we notice that the divergent terms are all contained in the case $k_{2}\rightarrow \infty$ and $k_{1}$ fixed, yielding
\begin{align}
B_{1}^{\Lambda}\equiv&\int\limits_{k_{1}k_{2}}\Delta^{2}(k_{1})\Delta(k_{2})\Delta(k_{2}-k_{1})\left[f_{0}^{\;(k_{1},\;p_{1})}+{\bar{f}}^{\;(k_{1},\;p_{1})}\right]\left[f_{0}^{\;(k_{1},\;p_{2})}+{\bar{f}}^{\;(k_{1},\;p_{2})}\right]\nonumber\\=&\int\limits_{k_{1}k_{2}}\Delta^{2}(k_{1})\Delta(k_{2})\Delta(k_{2}-k_{1})\Delta(k_{1}-p_{1})\Delta(k_{1}-p_{2})\nonumber\\
=&\quad\bar{B}_{1}^{\Lambda}+\beta_{1}^{\Lambda},\nonumber\\
\bar{B}_{1}^{\Lambda}\equiv&\int\limits_{k_{1}}\Delta(k_{1})\Delta(k_{1}-p_{1})\Delta(k_{1}-p_{2})\left[-\frac{I_{\text{log}}}{3}(\lambda^2)\right],\nonumber\\\beta_{1}^{\Lambda}\equiv&\int\limits_{k_{1}}\Delta(k_{1})\Delta(k_{1}-p_{1})\Delta(k_{1}-p_{2})\left[
\frac{b_{6}}{3}\ln\left(-\frac{k_{1}^{2}-\mu^{2}}{\lambda^2}\right)-\frac{8b_{6}}{9}\right].
\end{align}

Repeating the procedure in $\beta_{1}^{\Lambda}$ yields
\begin{align}
\bar{\beta}_{1}^{\Lambda}\equiv\int\limits_{k_{1}}\Delta(k_{1})\prod\limits_{i=1}^{2}f_{0}^{\;(k_{1},\;p_{i})}\left[\frac{b_{6}}{3}\ln\left(-\frac{k_{1}^{2}-\mu^{2}}{\lambda^2}\right)-\frac{8b_{6}}{9}\right]
=\frac{b_{6}}{3}\left[I_{\text{log}}^{(2)}(\lambda^{2})+
-\frac{8}{3}I_{\text{log}}(\lambda^{2})\right]
\end{align}
\noindent
Thus, the divergent content of $\Lambda_{B}^{(2)}$ is given by
\begin{align}
\Lambda_{B}^{(2)}\Big|_{\text{div}}\equiv\frac{-ig^{5}}{2}(\bar{\beta}_{1}^{\Lambda}+\bar{B}_{1}^{\Lambda}).
\end{align}

\begin{figure}[h!]
\begin{center} 
\includegraphics{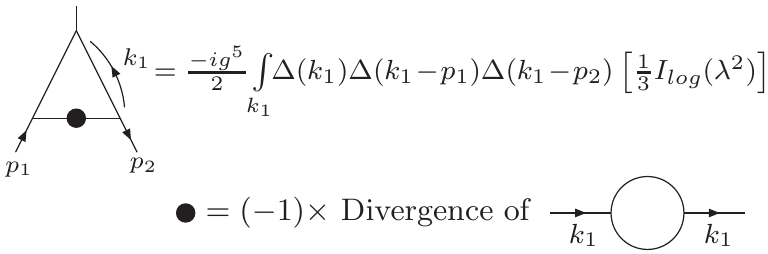}
\end{center}
\caption{Counterterm for $V_{B}^{(2)}$}
\label{ct:vb2}
\end{figure}

The last term is going to be removed after applying Bogoliubov's recursion formula since the counterterm for this graph is giving in fig. \ref{ct:vb2}.

After removing the subdivergence we obtain
\begin{align}
\frac{\bar{\Lambda}_{B}^{(2)}}{-ig^{5}}\equiv\frac{b_{6}}{6}I_{\text{log}}^{(2)}(\lambda^2)
-\frac{4b_{6}}{9}I_{\text{log}}(\lambda^2)+\mbox{finite}.
\end{align} 

Finally, we evaluate graph $V_{C}^{(2)}$. Denoting its amplitude as $\Lambda_{C}^{(2)}$ we have
\begin{align}
\frac{\Lambda_{C}^{(2)}}{-ig^{5}}\equiv\frac{1}{2}\int\limits_{k_{1}k_{2}}\Delta(k_{1})\Delta(k_{2})\Delta(k_{1}-p_{1})\Delta(k_{2}-p_{2})
\Delta(k_{1}+k_{2}-p_{1})\Delta(k_{1}+k_{2}-p_{2}).
\end{align}

As usual, we choose $n^{(k_{1})}_{1}=n^{(k_{2})}_{2}=n^{(k_{1}+k_{2})}_{1}=n^{(k_{1}+k_{2})}_{2}=1$ to find that the only divergent term comes from the case $k_{1}\rightarrow \infty$ and $k_{2}\rightarrow \infty$ simultaneously yielding
\begin{align}
C_{1}^{\Lambda}&\equiv\int\limits_{k_{1}k_{2}}f_{0}^{\;(k_{1},\;p_{1})}f_{0}^{\;(k_{2},\;p_{2})}f_{0}^{\;(k_{1}+k_{2},\;p_{1})}f_{0}^{\;(k_{1}+k_{2},\;p_{2})}\nonumber\\&=\int\limits_{k_{1}k_{2}}\Delta^{2}(k_{1})\Delta^{2}(k_{2})\Delta^{2}(k_{1}+k_{2}).
\end{align}

After performing the integral over $k_{2}$, the rules of IREG can be applied to yield
\begin{align}
\bar{\Lambda}_{C}^{(2)}\equiv\Lambda_{C}^{(2)}=-ig^{5}b_{6}\left[I_{\text{log}}(\lambda^2)+\mbox{finite}\right].
\end{align}

Grouping all the results we obtain that the divergent part of the three-point function at two loop order is given by
\begin{align}
\bar{\Lambda}^{(2)}_{\text{div}}\equiv(\bar{\Lambda}^{(2)}_{A}+\bar{\Lambda}^{(2)}_{B}+\bar{\Lambda}^{(2)}_{C})_{\text{div}}=ig^{5}\left[\frac{5b_{6}}{2}I_{\text{log}}^{(2)}(\lambda^2)
-\frac{17b_{6}}{3}I_{\text{log}}(\lambda^2)\right].
\label{vert}
\end{align}

\subsubsection{Two-loop renormalization group functions}

In summary, we could obtain the one and two-loop counterterms to the propagator and vertex function in a minimal subtraction scheme as 
\begin{align}
\Xi_{\text{ct}} &= -i\frac{g^2}{6}I_{\text{log}}(\lambda^2)-\frac{g^4}{6}\Bigg[\frac{5b_{6}}{3}I_{\text{log}}^{(2)}(\lambda^2)-\frac{29b_{6}}{9}I_{\text{log}}(\lambda^2)\Bigg];\\
\Lambda_{\text{ct}}&=-ig^{2}I_{\text{log}}(\lambda^2)-g^{4}\left[\frac{5b_{6}}{2}I_{\text{log}}^{(2)}(\lambda^2)-\frac{17b_{6}}{3}I_{\text{log}}(\lambda^2)\right].
\end{align}

Given the counterterms, it is an easy task to obtain the renormalization group functions. For completeness, we present the usual definitions
\begin{align}
\phi_{o}\equiv Z_{\phi}^{\frac{1}{2}}\phi, \quad 
\quad &g_{o}\equiv Z_{g}g,\quad \Xi_{\text{ct}} \equiv Z_{\phi}-1, \quad \Lambda_{\text{ct}}\equiv Z_{g}Z_{\phi}^{\frac{3}{2}}-1,\nonumber\\
&\gamma\equiv \lambda\frac{\partial \ln Z_{\phi}}{\partial \lambda}, \quad \beta 
\equiv-g \lambda \frac{\partial \ln Z_{g}}{\partial \lambda},
\end{align}
\noindent
where $\lambda$ plays the role of the renormalization group scale in the context of IREG.  
We finally obtain
\begin{align}
\gamma&=\frac{g^2}{6(4\pi)^{3}}+\frac{13g^4}{216(4\pi)^{6}}
+O(g^{6}),\\
\beta&=-\frac{3g^3}{4(4\pi)^{3}}-\frac{125g^5}{144(4\pi)^{6}}
+O(g^{6}).
\end{align}
This result agrees with the one in the literature \cite{Macfarlane:1974vp}. A treatment for massive theories can be performed in a similar fashion, for further details we refer the reader to \cite{ADRIANO}. Before moving to gauge theories, we emphasize that the algorithm applied in this section can be generalized to arbitrary loop order. We provide in \cite{ADRIANO} examples at 4- and n-loop order in the context of the $\phi^{3}$ model.

\subsection{Gauge Theories}

In this section, we will apply our procedure to massless gauge theories up to two-loop order. Since the complexity and number of diagrams is far superior than the example we provided for the $\phi^{3}$ model, we will not present all the details. Our main aim will be to discuss the importance of defining a normal form (as stated in sec. \ref{sec:ireg}) and collect known results. We will also only be interested in the UV behavior, meaning that all results consider only off-shell external momenta, avoiding the appearance of IR divergences. 

We begin with QED at 1-loop level. The divergent diagrams stand for the radiative correction to the photon and electron propagators, as well as the vertex diagram. In the context of IREG, they have been computed in \cite{Battistel:thesis,gaugen}. We will comment on the photon and electron propagator in more detail, since they illustrate some interesting features, while for the vertex we will just quote the result.

Starting with the photon propagator in the massless limit, after a standard application of Feynman rules, we obtain the amplitude
\begin{equation}
    i\Pi_{\mu\nu}(p)=(-)(-ie)^{2}\int_{k}\Tr \Bigg\{ \gamma_{\mu} \dfrac{i}{(\slashed{k})} \gamma_{\nu} \dfrac{i}{(\slashed{k}-\slashed{p})} \Bigg\}.
\end{equation}

In the context of IREG, our first step is to perform Dirac algebra in order to obtain a normal form compatible with gauge symmetry \cite{BRUQUE, nosso paper QCD}. For pedagogical purposes, we will skip this step and discuss the consequences. Removing the Dirac trace outside the integral, the amplitude can be written as 
\begin{equation}
    i\Pi_{\mu\nu}(p)=(-e^{2})\Tr\Bigg\{\gamma_{\mu}\gamma_{\alpha}\gamma_{\nu}\gamma_{\beta}(  I_{\alpha\beta}-I_{\alpha}p_{\beta})\Bigg\}\quad\mbox{where}\quad
    I_{\alpha_{1}\;\cdots\;\alpha_{n}}=\int_{k}  \dfrac{k_{\alpha_{1}}\cdots k_{\alpha_{n}}}{k^{2}(k-p)^{2}}. 
\end{equation}





It is an easy task to apply the IREG rules to these integrals to obtain
\begin{equation}
     i\frac{\Pi_{\mu\nu}}{(-e^{2})}=\dfrac{4}{3}\Bigg[  I_{\text{log}}(\lambda^2)-b\ln \Bigg( -\dfrac{p^{2}}{\lambda^{2}}\Bigg) +\dfrac{5}{3}b \Bigg](g_{\mu\nu}p^{2}-p_{\mu}p_{\nu})+\frac{2b}{3}g_{\mu\nu}
     \label{offending}
\end{equation}

    

As can be clearly seen, the result is NOT gauge invariant. The offending term comes from dismissing the first step on defining a normal form. Actually, by performing the trace before applying the regularization rules, one would obtain a term such as 
\begin{equation}
\int_{k}  \dfrac{k^{2}}{k^{2}(k-p)^{2}}=\int_{k}\dfrac{1}{(k-p)^{2}}=\lim_{\mu^{2}\rightarrow 0}\int_{k}  \dfrac{1}{(k-p)^{2}-\mu^{2}}=\lim_{\mu^{2}\rightarrow 0} I_{\text{quad}}(\mu^{2})=0
\label{Nk2}
\end{equation}
while by performing the traces afterwards one obtains
\begin{align}
g^{\alpha\beta}\int_{k}  \dfrac{k_{\alpha}k_{\beta}}{k^{2}(k-p)^{2}}
&=g^{\alpha\beta}\left\{\left(\frac{p_{\alpha}p_{\beta}}{3}-\frac{g_{\alpha\beta}p^{2}}{12}\right)\left[I_{\text{log}}(\lambda^{2})- b \ln\left(-\frac{p^2}{\lambda^2}\right) + \frac{13b}{6}\right]-\frac{g_{\alpha\beta}b p^{2}}{24}\right\}\nonumber\\
&=-\frac{b p^{2}}{6}
\label{gab_kakb}
\end{align}
where we have already discarded $I_{\text{quad}}(\mu^{2})$. Notice that the difference in both results has its origin in the fact that symmetric integration is not allowed in divergent integrals \cite{BRUQUE}  in dimension specific methods such as IREG. More importantly, 
the difference among Eq.~(\ref{Nk2}) and Eq.~(\ref{gab_kakb}) is exactly the offending term in Eq.~(\ref{offending}), since the Dirac trace will contribute with a global factor of (-4). 

Therefore, after the correct application of the rules of sec. \ref{sec:ireg} one obtains


\begin{equation}
     i\Pi_{\mu\nu}=-\dfrac{4e^{2}}{3}\Bigg[  I_{\text{log}}(\lambda^2)-b\ln \Bigg( -\dfrac{p^{2}}{\lambda^{2}}\Bigg) +\dfrac{5}{3}b \Bigg](g_{\mu\nu}p^{2}-p_{\mu}p_{\nu})
\end{equation}
which is transverse as expected. 

We move to the electron self-energy whose amplitude (in Feynman gauge) is given by


\begin{equation}
    i\Sigma(p)=(-ie)^{2}\int_{k}\Bigg\{ \dfrac{i}{k^{2}} \gamma_{\nu} \dfrac{i}{ (\slashed{p}-\slashed{k})} \gamma^{\nu} \Bigg\}.
\end{equation}

The application of IREG to it is straightforward, remembering that one must first perform the Dirac algebra
\begin{equation}
    i\Sigma(p)=-2e^{2}\Bigg\{\int_{k} \dfrac{\slashed{p}-\slashed{k}}{k^{2}(k-p)^{2}}\Bigg\} .
\end{equation}

The interesting point to be noticed here will be the difference in the finite part in contrast to the dimensional regularization result. The reason boils down to the difference in the definition of the Dirac algebra in the different methods. For instance, in IREG, 
\begin{equation}
    \gamma^{\beta}\slashed{p}\gamma_{\beta}=-2\slashed{p},
\end{equation}
while in DReg
\begin{equation}
    \gamma^{\beta}\slashed{p}\gamma_{\beta}=(2-d)\gamma^{\mu}=-2(1-\epsilon)\slashed{p}.
\end{equation}

In the context of DReg, the evanescent part~\cite{ZURICH,FIRENZE} will hit poles in $\epsilon$ generating more finite terms in contrast to the result in IREG. Nevertheless, our result is identical to the one of Dimensional Reduction after reducing the $\epsilon$-scalar coupling to the standard gauge coupling \cite{ZURICH}. Thus, we obtain

\begin{equation}
    i\Sigma(p)=(e)^{2}\slashed{p}\Bigg\{  I_{\text{log}}(\lambda^2)-b\ln \Bigg( -\dfrac{p^{2}}{\lambda^{2}}\Bigg) +2b \Bigg\} .
\end{equation}

Finally, the vertex function can be evaluated in a similar manner, we just quote its divergent part
\begin{equation}
    i\Lambda_{\mu}=e^{3}\gamma_{\mu}I_{\text{log}}(\lambda^2).
\end{equation}




\subsubsection{One and two-loop  RG functions for QED and QCD}

Proceeding to the renormalization program, the defintios in QED are
\begin{align}
    \psi^{0}=\sqrt{Z_{2}}\psi, \quad A_{\mu}^{0}=\sqrt{Z_{3}}A_{\mu}, \quad e_{0}=Z_{e}e.
\end{align}
With this, it is straightforward to obtain the renormalization functions 
\begin{align}
   \Lambda_{ct} =Z_{1}-1, \quad
    \Sigma_{ct}=Z_{2}-1, \quad
    \Pi_{ct}=Z_{3}-1, \quad Z_{1}=Z_{2}, \quad
    Z_{3}=1+\dfrac{4}{3}ie^{2}I_{\text{log}}(\lambda^2)
\end{align}
and check that the Ward identities are satisfied.

For QCD, the analysis is analogous. The usual definitions are
\begin{align}
    A_{0\mu}^{a}&=Z_{3}^{1/2}A_{\mu}^{a}, \quad c_{0}^{a}=\tilde{Z}_{3}^{1/2}c^{a}, \quad \psi_{0}=Z_{2}^{1/2}\psi,\quad g_{0}=Z_{g}g,\nonumber\\
    Z_{1}&\equiv Z_{g}Z_{3}^{1/2} , \quad Z_{4}\equiv Z_{g}^{2}Z_{3}^{2} , \quad \tilde{Z}_{1}\equiv Z_{g}\tilde{Z}_{3}Z_{3}^{1/2} , \quad Z_{1F}\equiv Z_{g}Z_{2}Z_{3}^{1/2}.
\end{align}

Given the large number of diagrams, we will just present the renormalization functions at 1-loop order, and refer the reader to \cite{Sampaio:2005pc}
\begin{align}
    Z_{3}&=1-ig^{2}\Bigg[ \dfrac{5}{3}C_{2}(G)-\dfrac{4}{3}n_{f}C(r) \Bigg] I_{\text{log}}(\lambda^2),\quad
    Z_{2}=1+ig^{2}C_{2}(r)I_{\text{log}}(\lambda^2),\nonumber\\
    \tilde{Z}_{3}&=1-\dfrac{ig^{2}}{2}C_{2}(G)I_{\text{log}}(\lambda^2),\quad
    Z_{1F}=1+ig^{2}(C_{2}(G)-C_{2}(r))I_{\text{log}}(\lambda^2),\nonumber\\
    \tilde{Z}_{1}&=1+\dfrac{ig^{2}}{2}C_{2}(G)I_{\text{log}}(\lambda^2), \quad
    Z_{1}=1+ig^{2}\Bigg( -\dfrac{2}{3}C_{2}(G)+\dfrac{4}{3}C(r)n_{f} \Bigg) I_{\text{log}}(\lambda^2),\nonumber\\
    Z_{4}&=1+ig^{2}\Bigg( \dfrac{1}{3}C_{2}(G)+\dfrac{4}{3}C(r)n_{f} \Bigg) I_{\text{log}}(\lambda^2).
\end{align}

Once again, the Slavnov-Taylor identities can be easily checked
\begin{equation}
    \frac{Z_{1}}{Z_{3}}=\frac{\tilde{Z}_{1}}{\tilde{Z}_{3}}=\frac{Z_{1F}}{Z_{2}}=\frac{Z_{4}}{Z_{1}}=1+ig^{2}C_{2}(G)I_{\text{log}}(\lambda^{2})
\end{equation}

\begin{figure}[h!]
\centering
\includegraphics[]{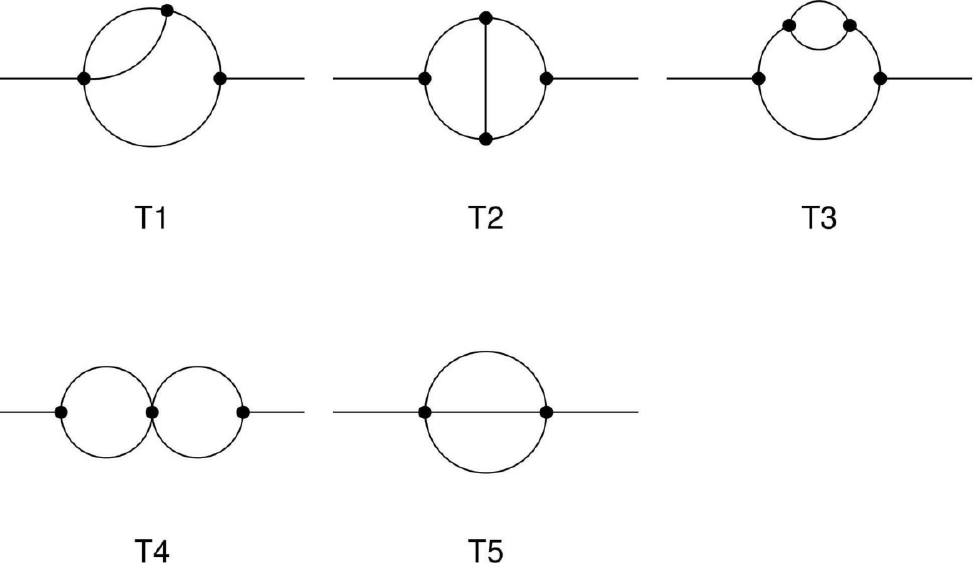}%
\caption{Two-loop topologies for two-point functions}
\label{Feynman}
\end{figure}

We move now to two-loop results. In the context of this review, we will focus on the calculation of the two-loop coefficient of the $\beta$-function gauge coupling both in QED and QCD. Since these coefficients are universal  \textbf{if} a renormalization substraction scheme independent of the mass is chosen (which is the case of IREG), we must recover the well-known results \cite{Vladimirov:1979zm,Larin:1993tp}. Given this objective, we will restrict ourselves to two-point functions only which amounts to the following topologies depicted in fig. \ref{Feynman}.

\begin{figure}[h!]
\centering
\includegraphics[width=0.20\textwidth]{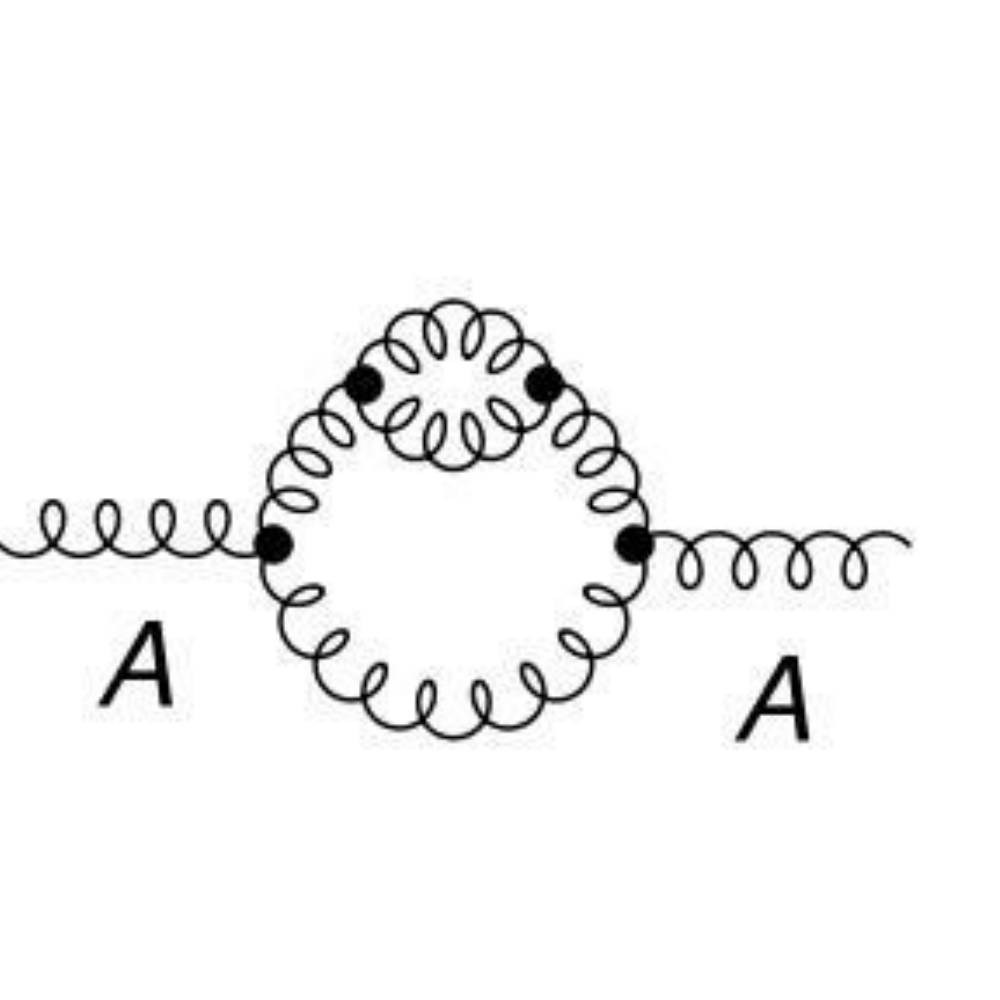}%
\caption{Two-loop diagram which contains a gluon loop as sub-diagram}
\label{fig:Feynman}
\end{figure}

The reason of this restriction is as follows: in the context of QED, the knowledge of the photon self-energy is enough to determine the gauge coupling $\beta$-function. In QCD, if one applies the background field method \cite{Abbott}, this statement is also true since $Z_{g}=Z_{A}^{-1/2}$. 
Notice that tadpoles diagrams were already removed, since they will amount to scaleless quadratic divergences (encoded as  $I_{\text{quad}}^{(l)}(\mu^2)$) which vanish in massless theories.
Given the general topologies shown in fig. \ref{Feynman}, we need to to fill them with the particle content of the theory under study. For instance, in QED, since four-point interactions do not occur, one can realize only topologies T2 and T3. This task was automatically performed by 
\textit{FeynArts} \cite{FeynArts}. 
After building the amplitudes in all theories, we designed in-house routines to perform Dirac and Lorentz algebra in 4-dimensions in the amplitude as a whole, making use of 
\textit{FormCalc} \cite{FormCalc}.
%
This is crucial to implement a normal form that respects gauge invariance at two-loop level as we briefly discuss. In order to illustrate this point, consider the diagram of Fig.
\ref{fig:Feynman}, 
\noindent
whose amplitude (in the Feynman gauge) is schematically given by
\begin{equation}
\mathcal{A} \propto \int_k \Pi_{\mu\nu\alpha\beta}(k,p)\int_l \frac{\mathcal{F}^{\alpha\beta}(q,k,p)}{q^2(q-k)^2},
\end{equation}
where we denote $q$ as the internal momentum of the sub diagram (gluon loop), $k$ is the internal momentum of the complete diagram, and $p$ the external momentum. As we presented in sec. \ref{sec:ireg}, one must first perform Dirac algebra, Lorentz contractions, etc, before applying the rules of IREG. Therefore, if  $\mathcal{F}^{\alpha\beta}$ contains a term like $q^{\alpha}q^{\beta}$ while $\Pi_{\mu\nu\alpha\beta}(k,p)$ has a term such as $g_{\alpha\beta}k_{\mu}k_{\nu}$, one will obtain
\begin{align}
\mathcal{B} \propto &\int_k \frac{k_{\mu}k_{\nu}}{k^4(k-p)^2}\int_q \frac{q^2}{q^2(q-k)^2}=
\int_k \frac{k_{\mu}k_{\nu}}{k^4(k-p)^2}\int_q \frac{1}{(q-k)^2}=0
\label{eq:b2}
\end{align}

Notice the similarity between the equation above and Eq.~(\ref{Nk2}). On the other hand, if one opts to perform the contraction with $g_{\alpha\beta}$ afterwards (see Eq.~\ref{gab_kakb}), 
\begin{align}
g_{\alpha\beta}\left[\int_q \frac{q^{\alpha}q^{\beta}}{q^2(q-k)^2}\right]=-\frac{bk^{2}}{6},
\end{align}
a non-null result to $\mathcal{B}$ will be obtained. As in the 1-loop case of QED, the second choice will result in the breaking of gauge invariance.


We proceed to a collection of results found in \cite{nosso paper QCD} in which we computed the two-loop correction to  $Z_{A}$, the renormalization function of the external gauge boson (the photon for QED, the background gluon field for QCD). Defining
\begin{align}
\label{eq:ZA pi}
Z_{A} = 1 + \frac{g^{2}}{(4\pi)^{2}} Z_{A}^{(1)} +  \frac{g^{4}}{(4\pi)^{4}}Z_{A}^{(2)}, 
\end{align}
one obtains for QED
\begin{align}
\label{eq:ZA1}
Z_{A}^{(1)}=-\frac{4}{3b}I_{\text{log}}(\lambda^{2}),\quad Z_{A}^{(2)}= - \frac{4}{b}I_{\text{log}}(\lambda^{2}),\\
\end{align}
and for QCD
\begin{align}
Z_{A}^{(1)}&=\left(\frac{11}{b}-\frac{2}{3b}n_{f}\right)I_{\text{log}}(\lambda^{2}),\\
Z_{A}^{(2)}&=\frac{54}{b^{2}}\left[I_{\text{log}}^{2}(\lambda^{2})-2bI_{\text{log}}^{(2)}(\lambda^{2})\right]+
\left(\frac{210}{b}-\frac{38}{3b}n_{f}\right)I_{\text{log}}(\lambda^{2}).
\label{eq:ZA2}
\end{align}
We have written the result for SU(3) group, while keeping the number of fermions as a free parameter $n_{f}$.
As standard, by defining the $\beta$-function by
\begin{equation}
\beta = -g\left[\beta_{0} \left(\frac{g}{4\pi}\right)^{2} + \beta_{1} \left(\frac{g}{4\pi}\right)^{4}\right];
\label{eq:beta pi}
\end{equation}  
one obtains
\begin{equation}
\beta_{0} = - \frac{1}{2}\lambda \frac{\partial}{\partial \lambda}Z_{A}^{(1)}, \quad \beta_{1} = - \frac{1}{2} \lambda \frac{\partial}{\partial \lambda}Z_{A}^{(2)}.
\label{eq:betai}
\end{equation}

Finally, using the results shown in eqs. \ref{eq:ZA1} to \ref{eq:ZA2}, we obtain the well-known one and two-loop contributions for the gauge $\beta$ coupling in QED and QCD \cite{Vladimirov:1979zm,Larin:1993tp}
\begin{align}
\text{QED}:\quad\quad &\beta_{0}= -\frac{4}{3}; \quad \quad\quad\;\,\beta_{1} = -4;\\
\text{QCD}:\quad\quad &\beta_{0} = 11-\frac{2}{3}n_{f}; \quad \beta_{1} = 102-\frac{38}{3}n_{f}.
\end{align}

Notice that the renormalization group $\beta$ function has been obtained in a gauge invariant fashion without explicitly evaluating the UV divergencies. This is in contrast with other schemes that operate in the physical dimension such as differential renormalization \cite{FREEDMAN} in which divergent expressions are replaced by finite. Moreover, in IREG the basic divergent objects depend explicitly on a renormalization scale $\lambda$ whereas the back-bones of ultraviolet divergencies in dimensional methods are terms proporcional to $1/\epsilon^n$, $n = 1, 2, \ldots  $,  as $\epsilon \rightarrow 0$.

.

\section{Concluding Remarks}

The purpose of this review was to present in a pedagogical way how the IREG method is put at work to comply with the powerful framework of BPHZ which is based on the fundamental principles of quantum field theory, unitarity, causality and locality . An algorithm has been shown that delivers the integrals involved in multi-loop amplitudes being decomposed in structures that are identified as the counterterms and divergencies of the order according to the BPHZ scheme. Various examples, ranging from the cubic scalar theory in six space time dimensions, to QED and QCD in the background field method have been worked out, highlighting the procedure. A further benefit of the method is that it delivers automatically all the necessary ingredients to obtain renormalization group functions, of which we have presented the beta functions to two-loop order of the above-mentioned theories, with known universal coefficients.

\vspace{6pt} 


\funding{~M.S. acknowledges a research grant from CNPq (Conselho Nacional de Desenvolvimento Cient\'\i fico e Tecnol\'ogico - 303482/2017-6). We acknowledge support from Fundação para a Ciência e Tecnologia (FCT) through the projects  UID/FIS/04564/2020 and  CERN/FIS-COM/0035/2019, and the networking support by the COST Action CA16201. This study was financed in part by the Coordenação de Aperfeiçoamento de Pessoal de Nível Superior – Brasil (CAPES) – Finance Code 001.
}

\end{paracol}
\reftitle{References}

\end{document}